\documentclass[11pt,a4paper]{amsart}
\usepackage{amsmath}
\usepackage{amssymb}
\usepackage{amsfonts}
\usepackage{hyperref}
\usepackage{amsthm}
\usepackage{mftinc}
\usepackage{graphicx}
\usepackage{epstopdf}

\def\={\ =\ }

\newcommand{\bea}{\begin{eqnarray}}
\newcommand{\eea}{\end{eqnarray}}
\newcommand{\beq}{\begin{equation}}
\newcommand{\eeq}{\end{equation}}
\newcommand{\bal}{\begin{align}}
\newcommand{\eal}{\end{align}}
\newcommand{\bit}{\begin{itemize}}
\newcommand{\eit}{\end{itemize}}

\newtheorem{theorem}{Theorem}
\theoremstyle{plain}

\newtheorem{proposition}{Proposition}

\numberwithin{equation}{section}
\input{tcilatex}

\begin{document}
\title{Torus knot polynomials and susy Wilson loops}
\author{Georgios Giasemidis}
\author{Miguel Tierz}
\address{\flushleft Rudolf Peierls Centre for Theoretical Physics\\
University of Oxford\\
1 Keble Road, Oxford OX1 3NP, UK}
\email{g.giasemidis1@physics.ox.ac.uk}
\address{\flushleft Departamento de An\'{a}lisis Matem\'{a}tico, Facultad de
Ciencias Matem\'{a}ticas \\
Universidad Complutense de Madrid \\
Plaza de Ciencias 3, Ciudad Universitaria, 28040 Madrid, Spain}
\email{tierz@mat.ucm.es}

\begin{abstract}
We give, using an explicit expression obtained in [V. Jones, Ann. of Math.
126, 335 (1987)], a basic hypergeometric representation of the HOMFLY
polynomial of $(n,m)$ torus knots, and present a number of equivalent
expressions, all related by Heine's transformations. Using this result the $%
(m,n)\leftrightarrow (n,m)$ symmetry and the leading polynomial at large $N$
are explicit. We show the latter to be the Wilson loop of 2d Yang-Mills
theory on the plane. In addition, after taking one winding to infinity, it
becomes the Wilson loop in the zero instanton sector of the 2d Yang-Mills
theory, which is known to give averages of Wilson loops in $\mathcal{N}$=4
SYM theory. We also give, using matrix models, an interpretation of the
HOMFLY polynomial and the corresponding Jones-Rosso representation in terms
of $q$-harmonic oscillators.
\end{abstract}

\maketitle

\section{Introduction}

The HOMFLY polynomial $X(\mathcal{K})$ is a knot invariant in the form of a
two-variable polynomial, generalizing the Jones polynomial \cite{Jones,
HOMFLY, PT}. It can be defined through a skein relation%
\begin{equation}
q^{N/2}X(L_{+})-q^{-N/2}X(L_{-})=(q^{1/2}-q^{-1/2})X(L_{0}),
\label{skein_relation}
\end{equation}%
where $L_{+}$, $L_{-}$ and $L_{0}$ are links formed by crossing and
smoothing changes on a local region of a link diagram with their direction
depicted as follows

\begin{center}
\includegraphics[scale=0.5]{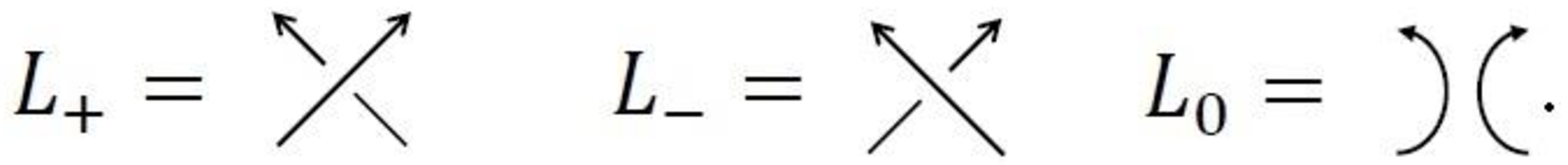}
\end{center}


In this work, we will examine the particular case of torus knots, which are
knots that lie on the surface of an unknotted torus in $\mathbb{R}^{3}$. A
torus knot, denoted either $T_{n,m}$ or $(n,m)$, where $n$ and $m$ are
coprime integers, has $n$ and $m$ windings along the non-contractible and
contractible cycles of a torus respectively.%

Our first task in this paper will be to analyze the HOMFLY polynomial $%
X(T_{n,m})$ for torus knots, using the explicit expression given by Jones 
\cite{Jones} and writing it in terms of terminating basic hypergeometric
series and $q$-orthogonal polynomials \cite{Gasper:1990,Koekoek}. For this,
we shall also use a random matrix model description, which originates in
Chern-Simons theory, and also discuss at the end of the paper implications
of the basic hypergeometric rewriting of the knot polynomial invariants in
gauge theory. For these reasons, let us also briefly review the basics
aspects of Chern-Simons theory, which provided a physical approach to Jones'
results \cite{Witten:1988hf}. Indeed, in the late 1980's Witten considered a
three dimensional gauge theory with a simply connected and compact
non-Abelian Lie group $G$ and the Chern-Simons action, which is given by 
\cite{Witten:1988hf}%
\begin{equation}
S_{\mathrm{CS}}(A)={\frac{k}{4\pi }}\int_{M}\mathrm{Tr}(A\wedge dA+{\frac{2}{%
3}}A\wedge A\wedge A),  \label{cs}
\end{equation}%
where $\mathrm{Tr}$ is the trace in the fundamental representation and $A$
is the connection, a 1-form valued on the corresponding Lie algebra, and $%
k\in \mathbb{Z}$ is the level. The manifold $M$ is a three dimensional
compact manifold which, in this work, will be chosen to be $S^{3}$. The $q$%
-parameter of Chern-Simons theory is defined in terms of the level $k$ by $%
q=\exp \left( 2\pi i/(k+N)\right) $.

In \cite{Witten:1988hf}, it was shown that Chern-Simons theory provides a
physical approach to three dimensional topology. In particular, the
observables of the gauge theory deliver three-manifold topological
invariants and knot polynomial invariants. In this way, in the Chern-Simons
theory approach, the HOMFLY polynomial \cite{HOMFLY,Jones,PT} of a knot $T$,
is given by the following Wilson loop average%
\begin{equation}
X(T)=\frac{q^{1/2}-q^{-1/2}}{c-c^{-1}}W_{\square }(T)  \label{homfly_1}
\end{equation}%
where $q$ is the usual Chern-Simons $q$-parameter, as above, $c=q^{N/2}$ and 
$W_{\mu }(T)$ denotes the normalized vacuum expectation value (vev) of the
Wilson loop in the representation $\mu $ which is defined as%
\begin{equation}
W_{\mu }(T):=\left\langle \mathrm{Tr}\,_{\mu }\left( P\exp \oint_{T}A\right)
\right\rangle =\langle \mathrm{Tr}_{\mu }\mathcal{U}_{T}\rangle ,
\label{wilson loop}
\end{equation}%
so the relationship (\ref{homfly_1}) involves the fundamental representation 
$\mu =\square $. 
The case of colored knot polynomials corresponds to a more general $\mu $,
and is a subject of much current interest.

The topological regularization of Wilson knot vevs requires a choice of
framing \cite{Atiyah}. This is essentially a choice of a companion of the
knot. This choice is parametrized by the number of times $f$ \ the companion
winds around the original knot. A change of framing by $f$ units leads to
multiplication by a phase%
\begin{equation*}
W_{\mu }(T)\rightarrow e^{2\pi ifh_{\mu }}W_{\mu }(T),
\end{equation*}%
where $h_{\mu }$ is the conformal weight of the corresponding
Wess-Zumino-Witten (WZW) primary field. The framing can also be specified by
adding \textit{ribbons} to the knot and thickening it into a band.

The case of $(n,m)$ torus knots 
can also be interpreted as a fractionally framed unknot with framing factor $%
f=m/n$. More precisely, it actually follows from the important result of
Rosso and Jones \cite{RJ} that the holonomy of creating a $(n,m)$ torus knot
is equivalent to the $n$-th power of the holonomy of a trivial knot,
together with a fractional framing $m/n$. In \cite{Brini:2012tk} and
previously in \cite{MM,Stevan:2010jh}, this is exploited to show that the
Wilson loop for the torus knot $T_{n,m}$ can be obtained by considering the
unknot $T_{n,1}$ while also acting with a fractional twist operator $\mathbb{%
T}^{m/n}$, introduced in \cite{MM}. %
%
The Jones-Rosso formula for torus knot is \cite{RJ,MM,Brini:2012tk}%
\begin{equation}
X(T_{n,m})=\frac{\left( q^{1/2}-q^{-1/2}\right) }{(c-c^{-1})}\sum_{\lambda
}c_{\square ,n}^{\lambda }q^{fC_{2}\left( \lambda \right) }\dim _{q}\lambda
\label{Rosso-Jones}
\end{equation}%
where $f=m/n$ and the coefficients $c_{\square ,n}^{\lambda }$ are only
non-zero when the sum is over Young tableaux of hook-shape. This simplified
expression, involving only hooks, is specific to torus knots and is due to
these knots being the closure of the braid word $\left( \sigma _{1}\sigma
_{2}...\sigma _{n-1}\right) ^{m}\in B_{n}$, where $\sigma _{i}$ are the
generators of the braid group $B_{n}$, in which case the braid, viewed as a
permutation, is just an $m$-cycle \cite{RJ}.

On the other hand, Chern-Simons theory on Seifert manifolds has a well-known
description in terms of matrix models \cite{Marino:2002fk,Tierz:2002jj}. In
this work we will use and focus on the unitary matrix model of $U(N)$
Chern-Simons theory on $S^{3}$ \cite{Okuda:2004mb,Romo:2011uc}%
\begin{equation}
Z_{N}=\frac{1}{N}\int_{\left( 0,2\pi \right] ^{N}}\prod_{j=1}^{N}\frac{d\phi
_{j}}{2\pi }\Theta _{3}({\,\mathrm{e}}\,^{i\phi _{j}}|q)\prod_{k<l}|{\,%
\mathrm{e}}\,^{i\phi _{k}}-{\,\mathrm{e}}\,^{i\phi _{l}}|^{2},  \label{RS}
\end{equation}%
where $\Theta _{3}({\,\mathrm{e}}\,^{i\phi }|q)$ denotes the Jacobi theta
function%
\begin{equation}
\Theta _{3}({\,\mathrm{e}}\,^{{\,\mathrm{i}\,}\phi }|q)=\sum_{n=-\infty
}^{\infty }\mathrm{q}^{n^{2}/2}\mathrm{e}^{in\phi }.  \label{weight}
\end{equation}%
This model can be solved exactly with orthogonal polynomials, the Rogers-Szeg%
\"{o} polynomials, which are the counterpart, on the unit circle \cite%
{Dolivet:2006ii}, of the Stieltjes-Wigert polynomials that solve the
Hermitian model \cite{Tierz:2002jj}. A more direct relationship between this
model and the trigonometric version of the original Chern-Simons matrix
model in \cite{Marino:2002fk} is given in \cite{Romo:2011uc}. We will be
interested in Wilson loop averages in the Rogers-Szeg\"{o} ensemble (\ref{RS}%
).

This paper is organized as follows. In Section \ref{Hypergeometric expr for
HOMFLY}, we give a basic hypergeometric expression for the HOMFLY polynomial
of a torus knot $(n,m)$ starting from the explicit result in Jones seminal
paper \cite{Jones}. 
Using several identities involving $q$-Pochhammer symbols and two different
Heine's transformations \cite{Gasper:1990}, we give various explicit basic
hypergeometric expressions, as well as the exact equivalence of the
different forms. A few conclusions from the hypergeometric expression, such
as making the $(m,n)\leftrightarrow $ $(n,m)$ symmetry invariance manifest,
are also drawn. %
We also explicitly show how the same result follows from the Jones-Rosso 
\cite{RJ} formula. Thus, in this Section we correct the hypergeometric
expressions given in \cite{Brini:2012tk}, show that their conjectured
expression follows directly from \cite{Jones} and prove that the different
hypergeometric representations available are all directly related through
Heine transformations.

In Section \ref{MM and SUSY Wilson loops}, we will show that these
expressions also follow from explicit computation, carried out long ago in 
\cite{Andrews:1984}, of a Wilson loop in the Rogers-Szeg{\"{o} ensemble (\ref%
{RS}). The average in the ensemble can be computed exactly either through
Rogers-Szeg\"{o} orthogonal polynomials or using a character expansion. We
show that the combination of both methods gives a new interpretation of both
the HOMFLY polynomial and the Jones-Rosso formula in terms of $q$-harmonic
oscillators. }

Finally, we shall show in the last Section that the large $N$ limit of the
HOMFLY polynomial, which is given by a Gauss hypergeometric function, is the
Wilson loop of 2d Yang-Mills theory on $S^{2}$ after decompactification and
that it also describes, in the limit $n\rightarrow \infty $, a family of BPS
Wilson loops. We conclude with a brief summary and avenues for further
research.

\section{Hypergeometric expressions for HOMFLY polynomials of torus knots}

\label{Hypergeometric expr for HOMFLY}In the last decade, the properties of
the (colored) Jones and HOMFLY polynomials as a $q$-holonomic system have
been studied in \cite{GL,G}. We will study a concrete realization of this
generic property by studying the different explicit basic hypergeometric
expressions for the HOMFLY polynomials of torus knots, correcting the
expressions in \cite{Brini:2012tk} and extending the description there by
using the theory of transformations of basic hypergeometric functions \cite%
{Gasper:1990}.

In the seminal work \cite{Jones}, we have an explicit computation of the
Jones-Ocneanu trace for $\left( \sigma_{1}\sigma _{2}...\sigma _{n-1}\right)
^{m}$ 
which accounts for the first explicit expression for the HOMFLY polynomial 

\begin{theorem}[{\protect\cite[Theorem 9.7]{Jones}}]
\label{JonesTheorem} Let $T$ be a torus knot of type $(n,m)$. Then the
HOMFLY polynomial is given by 
\begin{equation}
X(T_{n,m})=\left( \frac{1-q}{1-q^{n}}\right) \frac{q^{(N-1)(n-1)(m-1)/2}}{%
1-c^{2}}\sum\limits_{\substack{ \alpha ,\beta \geq 0  \\ \alpha +\beta =n-1}}%
(-1)^{\alpha }\frac{q^{\alpha m+\beta (\beta +1)/2}}{[\alpha ]_{q}![\beta
]_{q}!}\prod_{k=-\beta }^{\alpha }(q^{k}-c^{2}).  \label{X}
\end{equation}
\end{theorem}

The quantum numbers in \eqref{X} follow the convention 
$[n]_{q}=1-q^{n}$. Using Jones' expression \eqref{X} and, after simple
algebra, we find that%
\begin{equation}
X(T_{n,m})=(c^{2}q)^{(n-1)(m-1)/2}\frac{1-q^{-1}}{1-q^{-n}}\sum_{\beta
=0}^{n-1}q^{-m\beta }\left( \prod_{i=1}^{\beta }\frac{c^{2}q^{i}-1}{q^{i}-1}%
\right) \left( \prod_{j=1}^{n-1-\beta }\frac{c^{2}-q^{j}}{1-q^{j}}\right) .
\label{P}
\end{equation}%
which is equivalent to the expressions for the HOMFLY torus knots given in
the more recent work \cite{DGR} \footnote{%
Our notation and conventions differ from that of \cite{DGR}, where $q$ and $%
c $ are replaced by $q^{2}$ and $a$ respectively. In \cite{Jones} a $\lambda 
$ parameter ($\lambda q=c^{2})$ is used.}. 
Note that the initial steps to bring (\ref{P}) in basic hypergeometric form
have already been presented in \cite{Gorsky}. The next step is to write %
\eqref{P} 
in the form of a basic hypergeometric series, which is defined by \cite%
{Gasper:1990} 
\begin{equation}
_{2}\phi _{1}(b_{1},b_{2};c_{1};q,z)=\sum_{n=0}^{\infty }\frac{\left(
b_{1};q\right) _{n}\left( b_{2};q\right) _{n}}{\left( q;q\right) _{n}\left(
c_{1};q\right) _{n}}z^{n},  \label{basicdef}
\end{equation}%
where $\left( b;q\right) _{n}=$ $\prod\nolimits_{i=0}^{n-1}(1-bq^{i})$
denotes the $q$-Pochhammer symbol \cite{Gasper:1990}, see the Appendix \ref%
{BHF and qshifted factorials} for further details. After simple algebra we
obtain%
\begin{equation}
X(T_{n,m})=(c^{2}q)^{m(n-1)/2}\frac{[1]}{[n]}\sum_{\beta =0}^{n-1}q^{-m\beta
}\frac{\prod_{j=1}^{n-\beta -1}(q^{j/2}c^{-1}-cq^{-j/2})}{[n-\beta -1]!}%
\frac{\prod_{j=1}^{\beta }(q^{j/2}c-q^{-j/2}c^{-1})}{[\beta ]!}
\end{equation}%
where $[n]=q^{n/2}-q^{-n/2}$. 
We should now express the products above in terms of $(q^{A};q)_{\beta }.$
For this we use 
\begin{eqnarray}
\lbrack n-\beta -1]! &=&(-1)^{n-\beta -1}q^{-\sum_{k=1}^{n-\beta
-1}k/2}(q;q)_{n-\beta -1},  \notag \\
\prod_{j=1}^{n-\beta -1}(q^{j/2}c^{-1}-cq^{-j/2}) &=&(-1)^{n-\beta
-1}q^{-\sum_{k=1}^{n-\beta -1}k/2}c^{(n-\beta -1)}(q/c^{2};q)_{n-\beta -1}, 
\notag
\end{eqnarray}%
and the analogous expressions for the other two products. 
Next, we should express the terms $(-;q)_{n-\beta -1}$ in terms of $%
(-;q)_{\beta }$. This is possible by exploiting the identity \cite%
{Gasper:1990}%
\begin{equation}
\frac{(q/c^{2};q)_{n-\beta -1}}{(q;q)_{n-\beta -1}}=c^{2\beta }\frac{%
(q/c^{2};q)_{n-1}}{(q;q)_{n-1}}\frac{(q^{-(n-1)};q)_{\beta }}{%
(c^{2}q^{-(n-1)};q)_{\beta }}.
\end{equation}%
Finally, we find that%
\begin{equation}
X(T_{n,m})=(c^{2}q)^{(m+1)(n-1)/2}\frac{(q;q)_{1}(q/c^{2};q)_{n-1}}{(q;q)_{n}%
}{_{2}\phi _{1}}\left( c^{2}q,q^{-(n-1)};c^{2}q^{-(n-1)};q;q^{-m}\right) .
\end{equation}%
%
%
%
%
%
%
%
%
%
%
%
To show the equivalence between the different hypergeometric expressions, we
take into account the Heine's transformation \eqref{secondq}%
\begin{eqnarray}
{_{2}\phi _{1}}\left( c^{2}q,q^{-(n-1)};c^{2}q^{-(n-1)};q;q^{-m}\right) &=&%
\frac{(c^{2};q)_{\infty }(q^{1-n-m};q)_{\infty }}{(c^{2}q^{1-n};q)_{\infty
}(q^{-m};q)_{\infty }}\times  \notag \\
&&{_{2}\phi _{1}}\left( q^{1-m},q^{1-n};q^{1-n-m};q;c^{2}\right) .
\label{prev}
\end{eqnarray}%
We will see below that the basic hypergeometric function on the RHS of (\ref%
{prev}) is the one that appears, without derivation and after a proper
correction, in \cite{Brini:2012tk}. 
Let us check that the proportionality terms agree in the two expressions. To
do so, we first apply \eqref{I.5} and \eqref{I.2} 
and using \eqref{I.12} we simplify %
the pre-factor in the RHS of \eqref{prev} into 
\begin{equation}
\frac{(c^{2};q)_{\infty }(q^{1-n-m};q)_{\infty }}{(c^{2}q^{1-n};q)_{\infty
}(q^{-m};q)_{\infty }}=q^{(n-1)(1-m)/2}(1-q)c^{(m-1)(n-1)}\frac{(q;q)_{n+m-1}%
}{(q;q)_{n}(q;q)_{m}}.
\end{equation}%
%
%
%
%
Expressing it in terms of $[n]=q^{n/2}-q^{-n/2}$ we write
\begin{equation}
\frac{X(T_{n,m})}{q^{1/2}-q^{-1/2}}=c^{(m-1)(n-1)}\frac{[n+m-1]!}{[n]![m]!}{%
_{2}\phi _{1}}\left( q^{1-m},q^{1-n};q^{1-n-m};q;q^{N}\right) .  \label{PT}
\end{equation}%
This is exactly the expression that is also found below with a matrix model
computation. The appearance of the $\left( q^{1/2}-q^{-1/2}\right) $ factor
in the l.h.s. is due to the use of the unnormalized $q$-number $%
[n]=q^{n/2}-q^{-n/2}$. However, if we use the normalized $q$-number $\lfloor
n\rfloor =\left( q^{n/2}-q^{-n/2}\right) /\left( q^{1/2}-q^{-1/2}\right) $
we then have%
\begin{equation}
X(T_{n,m})=c^{(m-1)(n-1)}\frac{\lfloor n+m-1\rfloor !}{\lfloor n\rfloor
!\lfloor m\rfloor !}{_{2}\phi _{1}}\left(
q^{1-m},q^{1-n};q^{1-n-m};q;q^{N}\right) .  \label{PT2}
\end{equation}%
Notice that 
from the original (\ref{X}) (as was already pointed out in \cite{Jones}),
the symmetry $\left( m,n\right) \leftrightarrow (n,m)$ is not obvious at
all, whereas it is manifest from the expression (\ref{PT}) and the
definition of the basic hypergeometric function (\ref{basicdef}) which
implies an obvious $b_{1}\leftrightarrow b_{2}$ symmetry.

The basic hypergeometric functions obtained are, as expected, terminating
series since the first coefficient of the basic hypergeometric is $%
b_{1}=q^{1-m}$ \cite{Gasper:1990} and they can actually be written in terms
of the $q$-little Jacobi polynomials $p_{n}\left( a,b;q,x\right) ={_{2}\phi
_{1}}\left( q^{-n},abq^{n+1},aq;q,qx\right) $ \cite{Gasper:1990}, and hence
the HOMFLY polynomial can be written in terms of $p_{n}\left(
q^{-n-m},q^{-1};q,q^{N-1}\right) $. These polynomials also arise in a
natural way in the solution of the matrix model description of the HOMFLY
polynomials, when it is solved in terms of the eigenfunctions of the $q$%
-harmonic oscillator \cite{Andrews:1984}, as we shall see below.

Recall now that the HOMFLY polynomial of a knot $T$ is known to have the
following behavior \cite{LickMill, Correale:1994mx}%
\begin{equation}
X(T)=\sum_{i\geq 0}p_{i}(c^{2})\omega ^{2i},  \label{homfly_2}
\end{equation}%
where $\omega =q^{1/2}-q^{-1/2}$ and $c=q^{N/2}$. The polynomial $%
p_{0}(c^{2})$ is the leading term at large $N$ and has been of much interest
in the determination of the periodicity of knots \cite{Trac,Yok}. Notice
that taking the limit $q\rightarrow 1$ in (\ref{PT2}) \footnote{%
Note that, regarding the basic hypergeometric, we take its semiclassical
limit before specializing its variable $z$ with $c^{2}$. In the last Section
we will further interprete this limit and compare it with the $q\rightarrow
1 $ limit of the Chern-Simons matrix model.}, suggests the following
expression for the polynomial%
\begin{equation}
p_{0}(c^{2})=c^{(m-1)(n-1)}\frac{(m+n-1)!}{m!\ n!}{}%
_{2}F_{1}(1-m,1-n;1-m-n,c^{2}).  \label{p0}
\end{equation}%
This explicit expression is the one computed directly in \cite{Brini:2012tk}%
. 
In the last Section we will show that this is actually a Wilson loop in
two-dimensional Yang-Mills theory on $S^{2}$ after decompactification of the
sphere and we will compare this expression with the one that follows by
taking the full semiclassical limit of the unitary Chern-Simons matrix model.

\subsection{Equivalent hypergeometric expression and Heine's transformation}

\label{Hypergeometric and Heine}To complete our discussion on the HOMFLY
polynomial in terms of a basic hypergeometric function above, it is
interesting to further analyze the expression obtained in \cite{Brini:2012tk}%
, where it is shown that the Wilson loop of the framed unknot at winding
number $n$ with $f$ \ units of framing is given by%
\begin{equation}
\sum_{l=0}^{n}c^{2l+nf-n}(-1)^{n+l}\frac{1}{[n-l]![l]!}\frac{[nf+l-1]!}{%
[nf-n+l]!}  \label{U^m}
\end{equation}%
where the quantum number $n$ is defined as $[n]=q^{n/2}-q^{-n/2}$. 
Then, since $f\rightarrow m/n$ 
one has%
\begin{eqnarray}
&&\sum_{l=0}^{n}c^{2l+m-n}(-1)^{n+l}\frac{1}{[n-l]![l]!}\frac{[m+l-1]!}{%
[m-n+l]!}  \label{allgenus_2} \\
&=&\frac{(-1)^{n}c^{m-n}[m-1]!}{{[m-n]![n]!}}{}_{2}\phi _{1}\left(
q^{m},q^{-n};q^{1-n+m};q;q^{N+1}\right) .  \notag
\end{eqnarray}%
Notice that in \cite{Brini:2012tk} the expression given for the
hypergeometric is $_{2}\phi _{1}\left( m,-n;1-n+m;q;q^{N}\right) $ but the
correct expression is (\ref{allgenus_2}), taking into account that we both
use the standard definition (\ref{basicdef}).

In order to obtain the HOMFLY polynomial \eqref{homfly_1} from (\ref%
{allgenus_2}) one proceeds as above. First, we multiply (\ref{allgenus_2})
with $c^{-mn}$ in order to switch to the standard framing, then we shall
apply Heine's third transformation \eqref{thirdq}, 
and finally move to the standard convention for a torus knot, by changing $%
m\rightarrow -m$. Therefore, recalling (\ref{homfly_1}) we obtain%
\begin{eqnarray}
\frac{1}{q^{1/2}-q^{-1/2}}X(T_{n,m}) &=&\frac{c^{m-n-mn}}{c-c^{-1}}\frac{%
(-1)^{n}[m-1]!}{{[m-n]![n]!}}\frac{(q^{N};q)_{\infty }}{(q^{N+1};q)_{\infty }%
}{}_{2}\phi _{1}\left( q^{1+m},q^{1-n};q^{1-n+m};q;q^{N}\right)  \notag \\
&=&c^{(m+1)(1-n)}\frac{(-1)^{n+1}[m-1]!}{{[m-n]![n]!}}{}_{2}\phi _{1}\left(
q^{1+m},q^{1-n};q^{1-n+m};q;q^{N}\right) ,  \label{homfly_4_2}
\end{eqnarray}%
where in the last line, we have used that $\frac{(q^{N};q)_{\infty }}{%
(q^{N+1};q)_{\infty }}=(1-q^{N})=c(c^{-1}-c)$. Finally, switching to the
standard convention gives the corrected expression%
\begin{equation}
\frac{1}{q^{1/2}-q^{-1/2}}X(T_{n,m})=c^{(m-1)(n-1)}\frac{[n+m-1]!}{{[m]![n]!}%
}{}_{2}\phi _{1}\left( q^{1-m},q^{1-n};q^{1-n-m};q;q^{N}\right) ,
\label{H-last}
\end{equation}%
%
%
%
%
%
%
%
%
%
%
%
%
%
%
%
%
%
%
%
%
%
%
%
%
%
%
%
%
%
%
%
%
%
%
%
which coincides with the result \eqref{PT} obtained before, which followed
from \cite{Jones} or the equivalent expression in \cite{DGR}. Notice that
the expression (\ref{H-last}) is the one given in \cite{Brini:2012tk}, again
after the necessary correction pointed out below (\ref{allgenus_2}).
However, it is not derived there but simply given directly by the
straightforward promotion of the semiclassical result (\ref{p0}) to the $q$%
-deformed case, whereas the expression which is actually computed in \cite%
{Brini:2012tk} is (\ref{allgenus_2}). Thus, it was only conjectured in \cite%
{Brini:2012tk} and we have seen here that it follows from a Heine
transformation \cite{Gasper:1990} ((\ref{thirdq}) in the Appendix).

\subsection{Basic hypergeometric from the Jones-Rosso formula}

The Jones-Rosso formula for torus knots can be written as \cite{RJ} 
\begin{equation}
\bar{X}(T_{n,m})=\sum_{R_{n,s}}(-1)^{s}q^{fC_{2}(R_{n,s})/2}\mathrm{dim}%
_{q}(R_{n,s}),  \label{Um1f}
\end{equation}%
where $C_{2}$ denotes the quadratic Casimir and $R_{n,s}$ in \eqref{Um1f} is
a hook representation characterized by a Young tableaux $(n-s,1^{s})$. $\bar{%
X}(T)$ is the unnormalized HOMFLY polynomial, which satisfies 
\begin{equation}
\bar{X}(\mathrm{unknot})=\frac{c-c^{-1}}{q^{1/2}-q^{-1/2}}.
\end{equation}%
The Casimir for the hook representation $R_{n,s}=(n-s,1^{s})$ is given by 
\cite{Gross:1992tu} 
\begin{eqnarray}
C_{2}^{U(N)}(R_{n,s}) &=&Nn+(n-s)(n-s-1)-s^{2}-s  \notag \\
&=&Nn+n(n-1)-2ns,
\end{eqnarray}%
whereas the quantum dimensions for the same representations are given by 
%
%
%
\cite{Dolivet:2006ii}\footnote{%
The generic definition is 
\begin{equation}
\mathrm{dim}_{q}(R):=\prod_{x\in R}\frac{\lfloor N+c(x)\rfloor }{\lfloor
h(x)\rfloor }  \label{qdim_def_DT_final}
\end{equation}%
where $\lfloor n\rfloor =\frac{q^{n/2}-q^{-n/2}}{q^{1/2}-q^{-1/2}}$. %
\eqref{qdim_def_DT_final} reduces to \eqref{dimq_2} in the special case of
the Young diagram $R=R_{n,s}=(n-s,1^{s})$.}%
\begin{equation}
\mathrm{dim}_{q}(R_{n,s})=\frac{[N+n-s-1]!}{[n][n-s-1]![s]![N-s-1]!}.
\label{dimq_2}
\end{equation}%
We now show that \eqref{Um1f} can be written as a basic hypergeometric
function. We start by substituting the expressions for the quantum
dimensions and Casimir into \eqref{Um1f} 
\begin{equation}
\bar{X}(T_{n,m})=\sum_{s\geq 0}(-1)^{s}q^{fn(N+n-1-2s)/2}\frac{1}{[n]}\frac{%
[N+n-s-1]!}{[n-s-1]![s]![N-s-1]!}.
\end{equation}%
Recalling that $[n]=(q^{n/2}-q^{-n/2})$ and taking into account that $%
[n]!=(-1)^{n}q^{-\sum_{j=1}^{n}j/2}(q;q)_{n},$we have 
\begin{equation}
\bar{X}(T_{n,m})=\frac{q^{fn(N+n-1)/2+n(1-N)/2}}{(1-q^{n})}\sum_{s\geq
0}(-1)^{s}\frac{q^{-nfs+(s^{2}+s)/2}}{(q;q)_{s}}\frac{(q;q)_{N+n-s-1}}{%
(q;q)_{n-s-1}(q;q)_{N-s-1}}.
\end{equation}%
Applying the identity \eqref{I.10} to the terms $(q;q)_{N+n-s-1}$, $%
(q;q)_{n-s-1}$ and $(q;q)_{N-s-1}$ 
it follows that 
\begin{equation}
\bar{X}(T_{n,m})=\frac{q^{fn(N+n-1)/2+n(1-N)/2}}{(1-q^{n})}\frac{%
(q;q)_{N+n-1}}{(q;q)_{n-1}(q;q)_{N-1}}{_{2}}\phi
_{1}(q^{1-N},q^{1-n};q^{1-N-n};q,q^{-nf}).
\end{equation}%
Finally, we use the Heine's transformations \eqref{secondq} of $_{2}\phi
_{1} $ series 
which implies 
\begin{eqnarray}
\bar{X}(T_{n,m}) &=&q^{n(f-1)(N-1)/2+n^{2}f/2}\frac{(q;q)_{N+n-1}}{%
(q;q)_{n}(q;q)_{N-1}}\frac{(q^{-N};q)_{1-n}}{(q^{-nf};q)_{1-n}}  \notag \\
&&\times {_{2}}\phi _{1}(q^{1-nf},q^{1-n};q^{1-n-nf};q;q^{-N}).
\end{eqnarray}%
%
%
%
%
%
%
%
%
%
%
%
The prefactor can be further simplified using the identity \eqref{I.12}, 
thus 
\begin{eqnarray}
\bar{X}(T_{n,m}) &=&q^{n(f-1)(N-1)/2+n^{2}f/2}q^{(n-1)(N-fn)}(1-q^{N})\frac{%
(q;q)_{nf+n-1}}{(q;q)_{n}(q;q)_{nf}}  \notag \\
&&\times {_{2}}\phi _{1}(q^{1-nf},q^{1-n};q^{1-n-nf};q;q^{-N}).
\end{eqnarray}%
As a final step we want to relate this expression to \eqref{H-last} (also
(3.45) in \cite{Brini:2012tk}), after proper correction. First, we should
replace the $q$-Pochhammer symbols with the quantum number $%
[n]=q^{n/2}-q^{-n/2}$. Hence 
\begin{equation}
\bar{X}(T_{n,m})=(q^{N}-1)c^{nf+n-2}\frac{[nf+n-1]!}{[n]![nf]!}{_{2}}\phi
_{1}(q^{1-nf},q^{1-n};q^{1-n-nf};q;q^{-N}).  \label{TrUmk1f,final}
\end{equation}%
%
%
%
%
%
%
Now we first substitute $f\rightarrow m/n$, 
second we multiply with $c^{-mn}$ to switch to the standard framing and find%
\begin{equation*}
\bar{X}(T_{n,m})=(c-c^{-1})c^{(m-1)(1-n)}\frac{[m+n-1]!}{[n]![m]!}{_{2}}\phi
_{1}(q^{1-m},q^{1-n};q^{1-m-n};q;c^{-2}).
\end{equation*}%
Finally taking into account the transformation rule under mirror reflection 
\cite{Brini:2012tk} we obtain 
\begin{equation}
\bar{X}(T_{n,m}^{\ast })=(c-c^{-1})c^{(1-m)(1-n)}\frac{[m+n-1]!}{[n]![m]!}{%
_{2}}\phi _{1}(q^{1-m},q^{1-n};q^{1-m-n};q;c^{2}),  \label{barX-final}
\end{equation}%
where we used the fact that for $1-\alpha -\beta +\gamma =0$%
\begin{equation}
{_{2}}\phi _{1}(q^{-\alpha },q^{-\beta };q^{-\gamma };q^{-1};c)={_{2}}\phi
_{1}(q^{\alpha },q^{\beta };q^{\gamma };q;c).
\end{equation}%
We recall that the relationship between the normalized and unnormalized
HOMFLY polynomial should be given by%
\begin{equation}
\bar{X}(T)=\frac{c-c^{-1}}{q^{1/2}-q^{-1/2}}X(T)
\end{equation}%
and therefore, \eqref{barX-final} 
is exactly \eqref{H-last}. 

\section{Matrix model and SUSY Wilson loops}

\label{MM and SUSY Wilson loops}

The Wilson loop average $W_{n}(q;N)=\langle \mathrm{Tr}\,U^{n}\rangle _{%
\mathrm{RS}}$ in the Rogers-Szeg\"{o} ensemble corresponds to the Wilson
loop along the torus knot $T_{n,1}$, i.e. the unknot with winding number $n$%
. It is given explicitly by%
\begin{equation}
W_{n}(q;N)=\frac{1}{NZ_{N}}\int \prod_{j=1}^{N}\frac{d\phi _{j}}{2\pi }%
\Theta _{3}({\,\mathrm{e}}\,^{i\phi _{j}}|q)\sum_{m=1}^{N}{\,\mathrm{e}}%
\,^{in\phi _{m}}\prod_{k<l}|{\,\mathrm{e}}\,^{i\phi _{k}}-{\,\mathrm{e}}%
\,^{i\phi _{l}}|^{2}.  \label{Wilson}
\end{equation}%
Interestingly enough, this computation was carried out in full detail in the
early eighties by\ Andrews and Onofri \cite{Andrews:1984}, in the context of
lattice two-dimensional Yang-Mills theory, predating the appearance of pure
Chern-Simons theory. This is further explained in \cite{Romo:2011uc},
together with the relationship between the unitary matrix model and $q$%
-deformed 2d Yang-Mills theory \cite{Szabo:2013vva}. This specific model, in
contrast to the ones in 2d Yang-Mills theory with the heat-kernel or the
Wilson lattice action for example, is solvable both with the orthogonal
polynomials (Rogers-Szeg\"{o} polynomials) and with the character expansion
method employed in the case of the heat-kernel lattice action. Both
approaches are used in \cite{Andrews:1984} with identical results but
leading to equivalent yet complementary interpretations. The character
expansion computation is equivalent to the result carried out in \cite%
{Brini:2012tk} for the Hermitian ensemble, and gives rise to the Jones-Rosso
formula, whereas the orthogonal polynomial method leads to a new
interpretation of the observables in terms of $q$-harmonic oscillators on
the circle. Both computations are actually identical which is proven
rigorously in \cite{Andrews:1984} by establishing a non-trivial identity.

\subsection{Character expansion and $q$-deformed 2d Yang-Mills theory}

Let us now focus in the character expansion approach that solves (\ref%
{Wilson}), as shown in \cite{Andrews:1984}. This is directly related to the
results in \cite{Brini:2012tk}, however the character expansion of a theta
function is used instead of the Weyl character formula. This leads to an
interpretation in terms of the propagator of the $q$-deformed 2d Yang-Mills
theory, where the piece of the propagator that contains the Casimir of the
representation corresponds to the framing contribution discussed above.
Taking into account Frobenius formula, which relates the power-sum
polynomial with Schur polynomials and the fact that characters for full
cycles are always 1, 0 or -1, one has \cite{Andrews:1984}%
\begin{equation}
W_{n}(q;N)=\frac{1}{NZ_{N}}\int dU\det \left( \Theta _{3}({\,\mathrm{U}}%
|q)\right) \sum_{r\geq 0}(-1)^{r}\chi _{\left( n-r,1^{r}\right) }\left(
U\right) .  \label{Wchar}
\end{equation}%
Then, to give the explicit expression one needs the character expansion of $%
\det \left( \Theta _{3}({\,\mathrm{U}}|q)\right) $ which is worked out in 
\cite{Andrews:1984} and is actually a result previously known as Kostant
identity \cite{Kostant}, which is the character expansion for the theta
function $\gamma $ of a lattice $P$ \cite{Szabo:2013vva}%
\begin{equation}
\gamma \equiv \sum_{\lambda \in P}e^{\lambda }q^{\lambda ^{2}}=\sum_{\nu \in
P_{+}}q^{\left( \nu ,\nu +2\rho \right) }\left( \dim _{q}L_{\nu }\right)
\chi _{\nu },  \label{K}
\end{equation}%
where $P$ is the weight lattice of a simple and simply-laced Lie algebra $%
\mathfrak{g}$ and $\chi _{\nu }$ is the character of the irreducible
finite-dimensional module $L_{\nu }$ over $\mathfrak{g}$, and $\dim
_{q}L_{\nu }=\chi _{\nu }\left( q^{\rho }\right) $ is the quantum dimension.
Note that for $\mathfrak{g=u(n)}$ then $\gamma =\det \left( \Theta _{3}({\,%
\mathrm{U}}|q)\right) $. The expression (\ref{K}) has been often re-worked
(for example, in \cite{Andrews:1984}) and we can see that the r.h.s. of (\ref%
{K}) is actually the propagator of $q$-deformed 2d Yang-Mills theory \cite%
{Romo:2011uc,Szabo:2013vva}. The evaluation of (\ref{Wchar}) follows
immediately by the orthogonality of characters $\int dU\chi _{\lambda
}\left( U\right) \chi _{\mu }\left( U\right) =\delta _{\mu \lambda }$,
leading to%
\begin{eqnarray}
W_{n}(q;N) &=&\frac{1}{N}\sum_{r\geq 0}(-1)^{r}\int dU\sum_{\lambda
}q^{C_{2}\left( \lambda \right) }\dim _{q}\lambda \chi _{\lambda }\left(
U\right) \chi _{\left( n-r,1^{r}\right) }\left( U\right)  \notag \\
&=&\sum_{r\geq 0}(-1)^{r}q^{C_{2}\left( \left( n-r,1^{r}\right) \right)
}\dim _{q}\left( n-r,1^{r}\right) .  \label{Wchar2}
\end{eqnarray}%
We immediately have an interpretation of the Wilson loop average, and
therefore of the HOMFLY polynomial for torus knots, in terms of sums of
quantum dimensions of partitions with the shape of a hook, weighted with the
corresponding $q^{C_{2}\left( \lambda \right) }$ factors.

The orthogonal polynomials method \cite{Mehta} can also be applied to (\ref%
{Wilson}), giving another representation for the Wilson loop average%
\begin{equation}
W_{n}(q;N)=\frac{1}{N}\sum_{j=0}^{N-1}\langle \,\phi _{j}\,\rvert \,{}%
\widehat{z}^{n}\,\lvert \,\phi _{j}\,\rangle ,  \label{Wilson-h}
\end{equation}%
where $\phi _{j}(z)\,\ $denotes the orthonormal Rogers-Szeg\"{o} polynomials 
\cite{Koekoek}, which are characterized by%
\begin{equation}
\int \phi _{j}(z)\overline{\phi _{i}(z)}\Theta _{3}({\,\mathrm{z}}%
|q)dz=\delta _{i,j},  \label{orth}
\end{equation}%
and have the explicit expression%
\begin{equation*}
\phi _{j}(z)=\frac{1}{\left( q,q\right) _{j}^{1/2}}\sum_{k=0}^{j}{j\brack k}%
_{q}\left( -q\right) ^{\left( j-k\right) /2}z^{k},
\end{equation*}%
while $\widehat{z}$ in (\ref{Wilson-h}) denotes the angular position
operator $\widehat{z}\phi _{j}(z)=z\phi _{j}(z)$. These orthogonal
polynomials are the solution of the $q$-harmonic oscillator on the unit
circle \cite{MF} and, taking into account the Christoffel-Darboux identity 
\cite{Mehta}, the whole sum in (\ref{Wilson-h}) is not required and the
solution can be obtained just in terms of $\phi _{N}(z)$ and $\phi
_{N}^{\ast }(z)=z^{N}\phi _{N}(z^{-1})$ by constructing the density of
states of the matrix model%
\begin{equation}
\rho _{N}(\phi ;q)=\frac{1}{N}\Theta _{3}({\,\mathrm{e}}\,^{i\phi }|q)\left. 
\frac{\overline{\phi _{N}^{\ast }(\zeta )}\phi _{N}^{\ast }(z)-\overline{%
\phi _{N}(\zeta )}\phi _{N}(z)}{1-\zeta z}\right\vert _{z=\zeta ={\,\mathrm{e%
}}\,^{i\phi }},  \label{CD}
\end{equation}%
whose Fourier transform is the Wilson loop \cite{Andrews:1984} 
\begin{equation}
W_{n}(q;N)=\int d\phi {\mathrm{e}}\,^{in\phi }\rho _{N}(\phi ;q).
\label{intCD}
\end{equation}%
Thus, the two methods of solving the matrix model imply that 
\begin{eqnarray}
\sum_{r\geq 0}(-1)^{r}q^{C_{2}\left( \left( n-r,1^{r}\right) \right) }\dim
_{q}\left( n-r,1^{r}\right) &=&\frac{1}{N}\sum_{j=0}^{N-1}\langle \,\phi
_{j}\,\rvert \,{}\widehat{z}^{n}\,\lvert \,\phi _{j}\,\rangle  \notag \\
&=&\frac{(-1)^{n+1}q^{n/2}}{N}\frac{1}{1-q^{n}}{_{2}}\phi
_{1}(q^{n},q^{-n};q;q,q^{N+1})  \label{Andrews} \\
&=&\frac{(-1)^{n+1}q^{n/2}}{N}\frac{1-q^{N}}{1-q^{n}}{_{2}}\phi
_{1}(q^{1+n},q^{1-n};q;q,q^{N}).  \notag
\end{eqnarray}%
In addition, this is also directly proven, rigorously, as a consequence of a
very non-trivial identity established in \cite{Andrews:1984} (that we
collect in the Appendix \ref{BHF and qshifted factorials}).

As shown in \cite{Brini:2012tk} (see also previous computations in \cite%
{Tibi}), the consideration of the biorthogonal ensemble version of (\ref%
{Wilson}) modifies (\ref{Wchar2}) with a numerical factor $f=m/n$ in front
of $C_{2}\left( \left( n-r,1^{r}\right) \right) $. This is the fractional
framing and the resulting formula is the Jones-Rosso formula for torus
knots. Thus, the previous result extends immediately to the case of the
HOMFLY polynomials of torus knots and its Jones-Rosso formula representation
to 
\begin{equation}
X(T_{n,m})=\sum_{r\geq 0}(-1)^{r}q^{fC_{2}\left( \left( n-r,1^{r}\right)
\right) }\dim _{q}\left( n-r,1^{r}\right) =\frac{1}{N}\sum_{j=0}^{N-1}%
\langle \,\varphi _{j}\,\rvert \,{}\widehat{z}^{n}\,\lvert \,\widetilde{%
\varphi }_{j}\,\rangle ,  \label{knot}
\end{equation}%
where $\varphi _{j}$ and $\widetilde{\varphi }_{j}$ are biorthogonal
polynomials, defined by 
\begin{equation*}
\int \varphi _{n}(z)\Theta _{3}({\,\mathrm{z}}|q)z^{k}dz=\delta _{n,k}\quad 
\mathrm{and}\quad \int \widetilde{\varphi }_{n}(z)\Theta _{3}({\,\mathrm{z}}%
|q)z^{fk}dz=\delta _{n,k}.
\end{equation*}
This is due to the fact that the Jones-Rosso formula follows from the
biorthogonal version of the matrix model \cite{Brini:2012tk} and that the
spectral solution of the matrix model in terms of orthogonal polynomials (%
\ref{Wilson-h}) immediately generalizes to the biorthogonal case \cite%
{Dolivet:2006ii} (r.h.s. of \ref{knot}) by using the set of biorthogonal
polynomials $\varphi _{j}(z)$ and $\widetilde{\varphi }_{j}(z)$, which
generalizes (\ref{orth}), because for $f=1$ one has $\varphi _{n}(z)=%
\widetilde{\varphi }_{n}(z)=\phi _{n}(z)$.

Notice that, from what we have seen in this paper and the previous results
in \cite{Andrews:1984}, the basic hypergeometric expressions in (\ref%
{Andrews}) are computed explicitly from both the l.h.s. and r.h.s.
expressions in the first line, whereas in (\ref{knot}) have been computed
using the l.h.s. expression. It remains to use the biorthogonal polynomials
to do the explicit computation of the r.h.s. in (\ref{knot}). We shall
address this elsewhere together with their explicit realization of the $q$%
-harmonic oscillator with $q^{-f}$ with $f=m/n$.

\subsection{SUSY Wilson loops}

The matrix model expression for the Wilson loop (\ref{Wilson}) immediately
reminds of the corresponding average in a Hermitian matrix model with
Gaussian potential (GUE ensemble) 
\begin{equation}
\langle W_{\mu }\rangle =\frac{1}{\mathcal{Z}}\int \mathcal{D}M\,\frac{1}{N}%
\mathrm{Tr}_{\mu }\,e^{M}\,\exp \left( -\frac{2N}{\overline{\lambda }}%
\mbox{Tr}M^{2}\right) \,,  \label{G-mm}
\end{equation}%
where $M$ is an $N\times N$ Hermitian matrix and $\overline{\lambda }%
=g_{YM}^{2}N$ is a 't Hooft coupling. When $\mu $ is the antisymmetric
representation (Ferrers diagram is one column) then the trace is the one in (%
\ref{Wilson}). From the matrix model point of view the average (\ref{G-mm})
arises in the semiclassical, $q\rightarrow 1$, limit of (\ref{Wilson}),
since the unitary Chern-Simons matrix model becomes the GUE ensemble in this
limit \cite{Romo:2011uc}, since the unitary matrices are described through
their tangent space at the origin in such a limit \cite[Section 2]{Baik}.

The solution of (\ref{G-mm}) is then the same as the one in the previous
Section, but with the $q\rightarrow 1$ limit of the $q$-oscillators. Thus,
the matrix model average (\ref{G-mm}) is also computed exactly, with the
Hermite polynomials instead, giving the well-known result \cite%
{Drukker:2000rr}%
\begin{equation}
\langle W\rangle =\frac{1}{N}L_{N-1}^{(1)}\left( -\frac{\overline{\lambda }}{%
4N}\right) \exp \left[ \frac{\overline{\lambda }}{8N}\right] \,,
\label{2d-result}
\end{equation}%
where $L_{N-1}^{(1)}$ is a generalized Laguerre polynomial\footnote{%
The Rodrigues formula for the generalized Laguerre polynomials is $%
L_{n}^{(\alpha )}(x)=$ $\frac{x^{-\alpha }e^{x}}{n!}\frac{d^{n}}{dx^{n}}%
(e^{-x}x^{n+\alpha })$.}. Note that (\ref{2d-result}) differs from the pure
exponential behavior expected from exact 2d Yang-Mills theory. It is now
well-understood that (\ref{2d-result}) follows, after a rescaling of the
coupling constant, from only considering the zero-instanton sector of the 2d
Yang-Mills theory on $S^{2}$ and then taking the decompactification limit $%
R\rightarrow \infty $ where $R$ is the radius of $S^{2}$. This was explained
in \cite{Bassetto:1998sr} and the discussion there carries over to the study
of Wilson loops in $\mathcal{N}$=4 supersymmetric Yang-Mills (SYM) theory 
\cite%
{Drukker:2007qr,Drukker:2007yx,Drukker:2007dw,Giombi:2009ms,Bassetto:2008yf,Young:2008ed}
as we shall see in what follows.

The expression (\ref{G-mm}) was conjectured in \cite{Erickson:2000af}\cite%
{Drukker:2000rr} and proved in \cite{Pestun:2007rz} to describe 1/2-BPS
circular Wilson loops in $\mathcal{N}=4$ SYM theory. 
In \cite{Drukker:2007qr,Drukker:2007yx,Drukker:2007dw}, we also find the
analysis of 1/8-BPS Wilson loops. 
A remarkable result is that the Wilson loop expectation value is also given
by the following Gaussian matrix model average%
\begin{equation}
\langle W_{R}(\mathcal{C})\rangle _{4d}=\frac{1}{\mathcal{Z}}\int [dX]\,%
\mathrm{Tr}_{R}e^{X}\,e^{-\frac{\mathcal{A}^{2}}{2g_{YM}^{2}\mathcal{A}_{1}%
\mathcal{A}_{2}}\mathrm{Tr}\,X^{2}},  \label{W-vev}
\end{equation}%
where $\mathcal{A}_{1},\mathcal{A}_{2}$ are the areas singled out by the
Wilson loop and $\mathcal{A}=\mathcal{A}_{1}+\mathcal{A}_{2}=4\pi $. Notice
that this contains the original result involving 1/2-BPS circular Wilson
loops since, by simply taking $\mathcal{A}_{1}=\mathcal{A}_{2}=\mathcal{A}/2$%
, then the r.h.s is the Gaussian matrix model (\ref{G-mm}).

The character expansion (\ref{Wchar2}), with dimensions instead of quantum
dimensions, corresponds exactly to a Wilson loop in 2d Yang-Mills theory on $%
S^{2}$ with the heat-kernel lattice action. The Wilson loop average in this
case gives \cite{Andrews:1984}%
\begin{equation}
W_{n}^{(H.K.)}({\mathcal{\ A}_1},N)=\frac{e^{-\frac{g_{\mathrm{YM}}^{2}{%
\mathcal{A}_1}}{4}n(N+1-n)}}{nN}{}_{2}F_{1}(N+1,1-n;1+N-n,\mathrm{e}^{\frac{%
g_{\mathrm{YM}}^{2}\mathcal{A}_1}{2}\,n\,}).  \label{HK}
\end{equation}%
Indeed, for pure $U(N)$ Yang-Mills theory on a sphere $S^{2}$ with area $%
\mathcal{A}$ it holds \cite{Migdal,Rusakov}%
\begin{eqnarray}
W_{n}(\mathcal{A}-\mathcal{A}_1,\mathcal{A}_1)&=&{\frac{1}{\mathcal{Z}N}}%
\sum_{R,S}d_{R}\,d_{S}\exp \left[ -{\frac{g_{\mathrm{YM}}^{2}{(\mathcal{A}-%
\mathcal{A}_1)}}{4}}C_{2}(R)- {\frac{g_{\mathrm{YM}}^{2}{\mathcal{A}_1}}{4}}%
C_{2}(S)\right] \times  \notag \\
& &\times \int dU\mathrm{Tr}[U^{n}]\chi _{R}(U)\chi _{S}^{\dagger }(U),
\end{eqnarray}%
where $\mathcal{A}-\mathcal{A}_1$ and $\mathcal{A}_1$ are the areas singled
out by the Wilson loop. It is known that, in the decompactification limit $%
\mathcal{A}\rightarrow \infty $, ${\mathcal{A}_1}$ fixed, the following
expression is recovered \cite{Bassetto:1999dg,Bassetto:1999td}%
\begin{equation}
W_{n}({\mathcal{A}_1};N)=\frac{1}{nN}\,\exp \left( -\frac{g_{\mathrm{YM}}^{2}%
{\mathcal{A}_1}}{4}\,n(N+n-1)\right) \sum_{k=0}^{\infty }\frac{(-1)^{k}}{k!}%
\frac{\Gamma (N+n-k)}{\Gamma (N-k)\Gamma (n-k)}\mathrm{e}^{\frac{g_{\mathrm{%
YM}}^{2}{\mathcal{A}_1}}{2}\,n\,k}\,.  \label{series}
\end{equation}%
As happens with the expression for the HOMFLY polynomial, the series is
actually a finite sum, stopping at $k=n-1$ or $k=N-1$, depending on which
one is the smallest and, with the definition of the Gauss hypergeometric
function, (\ref{series}) can be immediately shown to give (\ref{HK}), which
now can be compared with (\ref{p0}).

Notice however the different specialization of the variable of the
hypergeometric function in (\ref{p0}) and (\ref{HK}); it depends on the
winding $n$ in (\ref{HK}) and on the rank $N$ in (\ref{p0}). However, we can
take into account the well-known rank-winding duality \cite%
{Bassetto:1999dg,Bassetto:1999td}%
\begin{equation*}
W_{n}^{(H.K.)}{}({\mathcal{A}_1},N)=W_{N}^{(H.K.)}{}\left( \frac{n}{N}{%
\mathcal{A}_1},n\right)
\end{equation*}%
which follows immediately from (\ref{HK}) and the manifest $%
_{2}F_{1}(a,b;c;z)=$ $_{2}F_{1}(b,a;c;z)$ property of the hypergeometric.
Thus, with the rescaled area $\widetilde{{\mathcal{A}}}=(n/N){\mathcal{A}_1}$%
, we also have%
\begin{equation}
W_{N}^{(H.K.)}{}(\widetilde{{\mathcal{A}}},n)=\frac{e^{-\frac{g_{\mathrm{YM}%
}^{2}\widetilde{{\mathcal{A}}}}{4}N(N-1+n)}}{nN}{}_{2}F_{1}(N+1,1-n;1+N-n,%
\mathrm{e}^{\frac{g_{\mathrm{YM}}^{2}\widetilde{{\mathcal{A}}}}{2}\,N\,}),
\label{HK2}
\end{equation}%
which is now of the form (\ref{p0}). Of course, the knot polynomial
invariant has two parameters $n$ and $m$, in addition to the $q$ and $N$,
whereas in the 2d Yang-Mills theory we just consider the winding $n$ of the
Wilson loop (the $q$ parameter is obviously identified with $\mathrm{e}^{g_{%
\mathrm{YM}}^{2}\widetilde{{\mathcal{A}}}/2\,\,}$). The sign difference
noticeable by comparing (\ref{p0}) with (\ref{HK2}) is simply due to the
convention chosen for the torus knots. The mirror image of the torus knot is
actually $T_{n,-m}$ \ 
and hence the knot polynomial for the mirror image of the torus knots gives
an exact correspondence between $p_{0}(c^{2})$ and $W_{N}^{(H.K.)}{}(%
\widetilde{{\mathcal{A}}},n)$, after specifying $m=N$ in the former.

Thus, we have seen explicitly that the polynomial $p_{0}(c^{2})$, the
leading term of the HOMFLY polynomial at large $N$ (\ref{homfly_2}), is the
Wilson loop of 2d Yang-Mills theory on $S^{2}$ after decompactification of
the sphere.

The respective large $n$ and large $N$ of (\ref{HK}) and (\ref{HK2}) reduces
the hypergeometric to a confluent hypergeometric \cite{Andrews:1984}, which
is precisely the modified Laguerre polynomial above, giving%
\begin{eqnarray*}
\lim_{\substack{ N\rightarrow \infty  \\ g_{\mathrm{YM}}^{2}N=\text{ fixed}}}%
W_{n}^{(H.K.)}{}({\mathcal{A}_1},N) &=&W_{n}{}({\mathcal{A}_1},N=\infty )=%
\frac{1}{n}L_{n-1}^{(1)}(\widehat{g}_{\mathrm{YM}}^{2}\mathcal{A}_1n/2)\exp
\left( -\frac{\widehat{g}_{\mathrm{YM}}^{2}\mathcal{A}_1n}{4}\right) , \\
\lim_{\substack{ n\rightarrow \infty  \\ n\mathcal{A}_1=\text{ fixed}}}%
W_{N}^{(H.K.)}{}(\widetilde{{\mathcal{A}}},n) &=&W_{N}{}(\widetilde{{%
\mathcal{A}}},n=\infty )=\frac{1}{N}L_{N-1}^{(1)}(g_{\mathrm{YM}}^{2}{%
\mathcal{A}_1}n/2)\exp \left( -\frac{g_{\mathrm{YM}}^{2}{\mathcal{A}_1}n}{4}%
\right) ,
\end{eqnarray*}%
with $\widehat{g}_{\mathrm{YM}}^{2}=g_{\mathrm{YM}}^{2}N$. The first
expression was computed long ago in \cite{KK, KK2, Ross}, whereas the second
is the one that appears in the perturbative resummation in the
Wu-Mandelstam-Leibbrandt prescription, which is equivalent to isolating the
zero-instanton contribution on $S^{2}$ (see \cite{Bassetto:1998sr} and
references therein).

Therefore, we find that the SUSY Wilson loop average is given by the large $%
N $ and large winding limit of the HOMFLY polynomial of a $\left( n,m\right) 
$ torus knot polynomial. In the following table we summarize the results,
including the corresponding expression in terms of a sum over Young
tableaux, which is the Jones-Rosso formula in the case of the HOMFLY
polynomial. In this way, summation in the third column is over Young
tableaux of hook shapes, denoted by $Y$ and $r$ is the length of the leg of
the hook \cite{MM} (the number of rows with one box)

\begin{center}
\begin{tabular}{|c|c|c|}
\hline
{\small Polynomial} & {\small Gauge theory} & {\small \textquotedblleft
Jones-Rosso'' form} \\ \hline
HOMFLY & $W_{n}${\small \ in $U(N)$\ CS theory on }$S^{3}$ & $%
\sum_{Y}(-1)^{r}q^{fC_{2}\left( Y\right) }\dim _{q}Y$ \\ \hline
$p_{0}\left( c^{2}\right) $ & $W_{n}$ {\small in $U(N)$\ 2d YM} & $%
\sum_{Y}(-1)^{r}q^{C_{2}\left( Y\right) }\dim Y$ \\ \hline
$p_{0}\left( c^{2}\right) ${\small \ with }$n\rightarrow \infty $ & $W_{N}$%
{\small \ in }$0${\small -instanton sector of 2d YM} & $\sum_{n\in 
\mathbb{N}
,Y}\frac{(-1)^{r}}{(2n)!}\dim Y_{2n}.$ \\ \hline
\end{tabular}
\end{center}

We emphasize that while the first and third lines in the table are given by
a Wilson loop in the Rogers-Szeg\"{o} and Gaussian matrix models,
respectively, the second one follows from just using dimensions, keeping the 
$\exp \left( -g_{YM}^{2}AC_{2}\left( \lambda \right) \right) $ term, which
gives the Wilson loop in the full 2d Yang-Mills theory. Interestingly, the
knot polynomial $p_{0}\left( c^{2}\right) $ can still be also related to the
BPS Wilson loop since taking one winding to infinity reduces the
hypergeometric to a confluent hypergeometric function $_{1}F_{1}\left(
1-n,2;n\lambda \right) $ \cite{Andrews:1984}, which is exactly the
hypergeometric representation of the modified Laguerre polynomial (\ref%
{2d-result}). The explicit expression in the third line follows from a
Taylor expansion of the $\mathrm{Tr}\,e^{M}$ term in the Gaussian matrix
model and the character expansion of the average of $\mathrm{Tr}M^{2n}$ over
a Gaussian Unitary ensemble \cite{Itzykson:1990zb}. The notation $Y_{2n}$
specifies explicitly that the number of boxes in the hook is $2n$. Thus, in
this case, we do not only sum over hooks of fixed size, but also over Young
tableaux of all (even) sizes. We will further discuss this result elsewhere.

\section{Outlook}

The results here suggest that an interesting line of further research could
be a systematic study of representations of knot and link polynomial
invariants in terms of the $q$-Askey scheme of $q$-orthogonal polynomials
and their corresponding expression in terms of a basic hypergeometric
function \cite{Koekoek}. From previous work, we know that the
Stieltjes-Wigert polynomials (or equivalently, the Rogers-Szeg\"{o}
polynomials), which are ${_{1}}\phi _{1}(q^{-n},0;q;-q^{n+1}x)$ basic
hypergeometric functions, at the bottom of the hierarchy \cite{Koekoek},
give the Witten-Reshetikhin-Turaev invariant on $S^{3}$ \cite{Tierz:2002jj}
(they were also used in \cite{Dolivet:2006ii} for computing quantum
dimensions). In what can be considered as a bottom-up approach through the $%
q $-Askey tableaux, we have seen in this work, using \cite{Andrews:1984},
how the HOMFLY polynomial of torus knots, which is a weighted sum of quantum
dimensions \cite{RJ}, requires the matrix element of $\widehat{z}^{m}$ in a $%
q$-oscillator basis (given by the Rogers-Szeg\"{o} polynomials) where $%
\widehat{z}$ is the operator $\widehat{z}\phi _{j}(z)=z\phi _{j}(z)$. These
matrix elements are given by the little $q$-Jacobi polynomials \cite%
{little,little2}, which are a step above in the $q$-Askey tableaux \cite%
{Koekoek}.

It would also be interesting to adopt a top-bottom approach, taking into
account the well-known result that shows that $q$-Racah polynomials, give
quantum invariants of links \cite{KR}. These polynomials descend through
various limits down to the Stieltjes-Wigert case \cite{Koekoek} passing
through the $q$-little Jacobi polynomials. The general result only has been
established in the case of the quantized universal enveloping algebra $%
\mathrm{U}_{q}(\mathfrak{sl}_{2})$ (Jones polynomial) \cite{KR} and the
generalization to $\mathrm{U}_{q}(\mathfrak{sl}_{N})$ (HOMFLY polynomial) is
not without difficulties \cite{Nawata:2013qpa}.

Other works in the recent literature also suggest that further
characterizations of the knot polynomial invariants in terms of basic
hypergeometric series can be expected. For example, the expressions in \cite%
{Arthamonov:2013rfa} for the colored HOMFLY polynomial of the Hopf link can
be brought into a basic hypergeometric form, as will be shown explicitly
elsewhere. Different results on knot polynomials and $q$-series can be found
in \cite{Andrews,Garoufalidis:2013rca}.

Another question of interest is to what extent trace averages in random
matrix ensembles can describe the Jones-Ocneanu trace of a representation of
a braid word. Recall that the latter is in general not a matrix trace, but a
weighted sum of matrix traces, since the Hecke algebra satisfies $\mathcal{H}%
_{n}(q)=\bigoplus\nolimits_{\lambda \vdash n}M_{\lambda }$ where each $%
M_{\lambda }$ is a two-sided ideal, isomorphic to a full matrix algebra over
the field $K$. The weights are known to be given by Schur polynomials and we
expect to show elsewhere that precisely the unitary matrix models that arise
in Chern-Simons theory on $S^{3}$ can be interpreted as integral
representations of Schur polynomials \cite{IP}.

Inspection of the corresponding Jones-Rosso formula for the colored HOMFLY
polynomial of torus knots and links \cite{Zhu} suggests that the matrix
model formalism can be definitely extended to that case, by consideration of
a more general trace average, such as $\left( \mathrm{Tr}U\right) ^{\alpha
_{1}}\left( \mathrm{Tr}U^{2}\right) ^{\alpha _{2}}...\left( \mathrm{Tr}%
U^{r}\right) ^{\alpha _{r}}$, in the unitary Chern-Simons matrix model. It
would be interesting to try to generalize any of the analytical methods
discussed here to this more general setting. The $q$-oscillator method using
the explicit expression of a power of the $\widehat{z}$ operator, in terms
of creation and annihilation operators, denoted by $\widehat{\mathfrak{a}}%
^{\dag }$ and $\widehat{\mathfrak{a}}$ respectively, of the $q$-harmonic
oscillator \cite{Andrews:1984}%
\begin{equation*}
\widehat{z}^{n}=\left( \frac{q+\widehat{\mathfrak{a}}^{\dag }}{1+q\widehat{%
\mathfrak{a}}}\right) ^{n}=\prod\limits_{j=1}^{n}\left( q^{2j-1}+\widehat{%
\mathfrak{a}}^{\dag }\right) \prod\limits_{k=0}^{\infty }\left( 1+q^{2k-1}%
\widehat{\mathfrak{a}}\right) ,
\end{equation*}%
seems an interesting possibility. There also exists the possibility of
computing knot polynomials through integrals, generalizing the one in (\ref%
{intCD}) for the HOMFLY polynomial of torus knots, of the diagonal
Christoffel-Darboux kernel (\ref{CD}) of $q$-harmonic oscillators. This
kernel, both in the biorthogonal and standard case, has been recently
further characterized in \cite{TK1,TK2}.

We have also seen that the consideration of different BPS Wilson loops in $%
\mathcal{N}$=4 theory is a subject of much current interest, with
applications also to the study of the radiation of a moving quark in $%
\mathcal{N}$=4 theory \cite{Correa} and in the study of entanglement
entropies \cite{LM}. The relationship shown here with the semiclassical
limit of Chern-Simons theory and with knot theory could be extended to more
general Wilson loops, which involve Gaussian averages of more general
traces, such as $\left( \mathrm{Tr}U\right) ^{\alpha _{1}}\left( \mathrm{Tr}%
U^{2}\right) ^{\alpha _{2}}...\left( \mathrm{Tr}U^{r}\right) ^{\alpha _{r}}$%
, and therefore may arise as semiclassical limits of colored HOMFLY
polynomials of torus knots.\bigskip

\section*{Acknowledgments}

We thank Mizan Rahman and Satoshi Nawata for valuable comments. GG would
like to thank the Rudolf Peierls Centre for Theoretical Physics, University
of Oxford for the kind hospitality during the completion of this work. MT
acknowledges financial support from a Juan de la Cierva Fellowship, from
MINECO (grant MTM2011-26912) and the European CHIST-ERA project CQC (funded
partially by MINECO grant PRI-PIMCHI-2011-1071). MT thanks J{\o }rgen
Andersen for the warm hospitality and for discussions at the Nielsen retreat
of the Centre for Quantum Geometry of Moduli Spaces, Aarhus University,
Denmark.


\appendix

\section{Basic hypergeometric functions, $q$-shifted factorials identities
and quantum dimension}

\label{BHF and qshifted factorials} The hypergeometric function $%
_{2}F_{1}(a,b;c;z)$ was introduced by Gauss in 1812. It is defined by 
\begin{equation*}
{}_{2}F_{1}(a,b;c;z)=\sum_{n=0}^{\infty }\frac{\left( a\right) _{n}\left(
b\right) _{n}}{\left( c\right) _{n}}z^{n},
\end{equation*}%
where $\left( a\right) _{n}$ denotes the shifted factorial, given by $\left(
a\right) _{0}=1$ and $\left( a\right) _{n}=\Gamma \left( n+a\right) /\Gamma
\left( a\right) $ for $n=1,2,...$. The basic hypergeometric $_{2}\phi _{1}$%
-series was introduced in 1846 by Heine and it is given by \cite{Gasper:1990}
\begin{equation}
_{2}\phi _{1}(a,b;c;q,z)=\sum_{n=0}^{\infty }\frac{\left( a;q\right)
_{n}\left( b;q\right) _{n}}{\left( q;q\right) _{n}\left( c;q\right) _{n}}%
z^{n},  \label{H}
\end{equation}%
where the $q$-shifted factorial is defined as \cite[Appendix I]{Gasper:1990} 
\begin{equation}
(a;q)_{n}=%
\begin{cases}
1, & n=0, \\ 
\prod_{k=0}^{n-1}(1-aq^{k}), & n>0, \\ 
\prod_{k=1}^{n}\frac{1}{(1-aq^{-k})}, & n<0.%
\end{cases}%
\end{equation}%
and it is assumed that $c\neq q^{-m}$ for $m=0,1,...$. The following
identities of the $q$-shifted factorial hold \cite[Appendix I]{Gasper:1990} 
\begin{eqnarray}
(a;q)_{-n} &=&\frac{1}{(aq^{-n};q)_{n}}=\frac{(-q/a)^{n}}{(q/a;q)}q^{\binom{n%
}{2}} ,  \label{I.2} \\
(a;q)_{n} &=&\frac{(a;q)_{\infty }}{(aq^{n};q)_{\infty }},  \label{I.5} \\
(a;q)_{n-k} &=&\frac{(a;q)_{n}}{(q^{1-n}/a;q)_{k}}\left( -\frac{q}{a}\right)
^{k}q^{{\binom{k}{2}}-nk},  \label{I.10} \\
(q^{-n};q)_{k} &=&\frac{(q;q)_{n}}{(q;q)_{n-k}}(-1)^{k}q^{{\binom{k}{2}}-nk},
\label{I.12}
\end{eqnarray}%
where ${\binom{n}{2}}=n(n-1)/2$. 
The transformation by Euler for the $_{2}F_{1}(a,b;c;z)$ hypergeometric
function is \cite{Morse:1953} 
\begin{eqnarray}
{}_{2}F_{1}(a,b;c;z) &=&(1-z)^{-a}{}_{2}F_{1}(a,c-b;c;\frac{z}{z-1})
\label{first} \\
&=&(1-z)^{-b}{}_{2}F_{1}(c-a,b;c;\frac{z}{z-1})  \label{second} \\
&=&(1-z)^{c-a-b}{}_{2}F_{1}(c-a,c-b;c;z).  \label{third}
\end{eqnarray}%
For the basic (or $q$-) hypergeometric series we have the Heine's
transformations.

\begin{proposition}
The basic hypergeometric function (\ref{H}) satisfies \cite[Appendix III]%
{Gasper:1990}, 
\begin{eqnarray}
{}_{2}\phi _{1}(a,b;c;q,z) &=&\frac{(b;q)_{\infty }(az;q)_{\infty }}{%
(c;q)_{\infty }(z;q)_{\infty }}{}_{2}\phi _{1}(c/b,z;az;q,b)  \label{firstq}
\\
&=&\frac{(c/b;q)_{\infty }(bz;q)_{\infty }}{(c;q)_{\infty }(z;q)_{\infty }}{}%
_{2}\phi _{1}(abz/c,b;bz;q,c/b)  \label{secondq} \\
&=&\frac{(abz/c;q)_{\infty }}{(z;q)_{\infty }}{}_{2}\phi
_{1}(c/a,c/b;c;q,abz/c).  \label{thirdq}
\end{eqnarray}%
\appendix
\end{proposition}

Regarding identities, a very non-trivial one was proven by Andrews and
Onofri, which demonstrates the equivalence between the character expansion
and the orthogonal polynomial solution \cite{Andrews:1984}:%
\begin{eqnarray}
\sum_{j\geq 0}\frac{(q^{s};q)_{j}(q^{-s};q)_{j}q^{(N+1)j}}{(q;q)_{j}^{2}} &=&%
\frac{(-1)^{s-1}(1-q^{s})}{(q;q)_{N}}\sum_{r=0}^{N}(-1)^{r}(N-r)(1-q^{N-r}) 
\notag \\
&&\times \sum_{n=0}^{r}{N\brack n}_{q}{N\brack r-n}_{q}q^{n+\binom{2n-r+s}{2}%
}  \label{AO_expr}
\end{eqnarray}%
where ${A\brack B}_{q}=\frac{[A]_{q}!}{[A-B]_{q}![B]_{q}!}$.



\end{document}